\UseRawInputEncoding
\documentclass[12pt]{iopart}

\usepackage{graphicx}

\begin{document}

\title[Measurement of the $^{91}$Zr(p,$\gamma$)$^{92m}$Nb cross section]{Measurement of the $^{91}$Zr(p,$\gamma$)$^{92m}$Nb cross section motivated by type Ia supernova nucleosynthesis}

\author{Gy~Gy\"urky, Z~Hal\'asz, G~G~Kiss, T~Sz\"ucs, R Husz\'ank, Zs~T\"or\"ok, Zs~F\"ul\"op}
\address{Institute for Nuclear Research (ATOMKI), Debrecen, Hungary}
\ead{gyurky@atomki.hu}
\author{T Rauscher}
\address{Department of Physics, University of Basel, Klingelbergstrasse 82, CH-4056 Basel, Switzerland}
\address{Centre for Astrophysics Research, University of Hertfordshire, College Lane, Hatfield AL10 9AB, United Kingdom}
\author{C Travaglio}
\address{INFN - Istituto Nazionale Fisica Nucleare, Turin, Italy}


\begin{abstract}
The synthesis of heavy, proton rich isotopes is a poorly understood astrophysical process. Thermonuclear (type Ia) supernova 
explosions are among the suggested sites and the abundance of some isotopes present in the early solar system may be used to test 
the models. $^{92}$Nb is such an isotope and one of the reactions playing a role in its synthesis is $^{91}$Zr(p,$\gamma$)$^{92}$Nb. 
As no experimental cross sections were available for this reaction so far, nucleosynthesis models had to solely rely on theoretical 
calculations. In the present work the cross section of $^{91}$Zr(p,$\gamma$)$^{92m}$Nb has been measured at astrophysical 
energies by activation. The results excellently confirm the predictions of cross sections and reaction rates for 
$^{91}$Zr(p,$\gamma$)$^{92}$Nb, as used in astrophysical simulations. 
\end{abstract}

%
\vspace{2pc}
\submitto{\JPG}
%
%
%

\section{\label{sec:intro}Introduction}

The proton rich stable isotopes of chemical elements heavier than iron represent a special category in nuclear astrophysics. As 
opposed to the more neutron rich species, these so-called p-nuclei between about $^{74}$Se and $^{196}$Hg are not synthesized via 
neutron capture reactions in the astrophysical s- and r-processes \cite{RevModPhys.83.157,RevModPhys.93.015002}. Several possible 
astrophysical sites and mechanisms (like $\gamma$-, rp-, $\nu$- or $\nu$p-processes) are considered which could contribute to their 
production, summarized under the name p-process \cite{ARNOULD20031,Rauscher_2013}. 

Processes in different astrophysical conditions are indeed needed as no single model has been found yet which could reproduce the 
abundances of all the p-isotopes observed in the solar system. Thermonuclear (type Ia \cite{Travaglio2014}) as well as core 
collapse (CCSN \cite{Travaglio_2018}) supernovae have been extensively studied as possible sites for the p-process. Nevertheless, 
the reproduction of the observed p-isotope abundances by the models remains poor, especially in some specific mass regions. One particularly 
important mass region is around $^{92}$Mo where the isotopes
are notoriously underproduced by the models. 

$^{92}$Nb, the long-lived (t$_{1/2}$\,=\,3.47$\cdot 10^7$ years) radioactive isobar of $^{92}$Mo has additional importance as its abundance at the time of the solar system formation can be inferred from the isotopic ratio measurement of primitive meteorites \cite{Habae2017750118}. In 1996 Harper \cite{Harper1996} found first evidence for live $^{92}$Nb in the early Solar System material by measuring a small excess of $^{92}$Zr in rutile (TiO$_2$) extracted from an iron meteorite. Different studies of explosive nucleosynthesis tried to explain the observed abundance of meteoritic $^{92}$Nb analyzing the possible astrophysical sites in core collapse or thermonuclear supernovae. Despite many years of studies, its origin is still uncertain. 

Whereas $^{93}$Nb is about 85\% $s$-process and 15\% $r$-process (from the original predictions by Arlandini et al. 
\cite{Arlandini1999} and Travaglio et al. \cite{Travaglio1999}), $^{92}$Nb is an important isotope since it is produced by the 
$\gamma$-process but is completely shielded from contributions from rp-or $\nu$p-processes \cite{Dauphas2003}. For this reason it 
can be particularly helpful to test models of p-process nucleosynthesis. $^{92}$Nb is usually normalized to $^{92}$Mo because both 
are p-process nuclides while $^{93}$Nb has an $s$-process origin (by the radiogenic decay of $^{93}$Zr). The underproduction of 
$^{92}$Mo in the $\gamma$-process could, in principle, be compensated by contributions of the rp-or $\nu$p-processes but this would 
lead to a too low $^{92}$Nb/$^{92}$Mo ratio at Solar System birth (as discussed in \cite{Dauphas2003}). 

In SNIa models, theoretical estimates for the ratio $^{92}$Nb/$^{92}$Mo have been presented by Travaglio et al. \cite{Travaglio2014}. These authors analyzed in detail the production of $^{92}$Nb (and also the other radioactive p-nucleus $^{146}$Sm) in SNIa using multidimensional models and concluded that such an origin is plausible for both radionuclides.

More recently, Nishimura et al. \cite{Nishimura2018}, investigating the same model presented by \cite{Travaglio2014}, found that the uncertainties stemming from uncertainties in the astrophysical reaction rates are small compared to the uncertainties arising from the choice of site, explosion model, and numerical treatment of the explosion hydrodynamics, giving more strength to the first result published in 2014 \cite{Travaglio2014}.

Regarding CCSNe, the first analysis of the possible origin of $^{92}$Nb in CCSNe has been presented by \cite{Woosley1978}. These authors concluded that $^{92}$Nb can possibly form in these stars {\it within experimental error}, referring to uncertainties in the measurements of the ratio $^{92}$Nb/$^{93}$Nb extracted from the Earth as well as to the estimated lifetime uncertainties of $^{92}$Nb.

This old work has been updated by different authors over the years, see \cite{Rauscher2002} for a review. Rauscher et al. \cite{Rauscher2002} also demonstrated that with their models they can reproduce the Solar System $^{92}$Nb/$^{92}$Mo ratio but, at the same time, they underproduce the amount of $^{92}$Mo present in cosmic abundances. They concluded that CCSNe are unlikely contributors to p-process nuclides in the Mo--Ru mass region.

More recently, Hayakawa et al. \cite{Hayakawa2013} presented new calculations in CCSNe which demonstrated a novel origin for $^{92}$Nb via neutrino-induced reactions. Their calculations showed that the observed ratio of $^{92}$Nb/$^{93}$Nb $\approx$ 10$^{-5}$ can be explained by this process. Nevertheless, the authors did not consider the production of the others Mo--Ru isotopes (neither the other p-isotopes) in the same process/stellar source.

For an updated discussion about the possible stellar sources of $^{92}$Nb (and $^{146}$Sm) in SNIa and/or CCSNe, see the overview of Lugaro et al. \cite{Lugaro2016}.  

The above summarized importance of $^{92}$Nb necessitates the good nuclear physics knowledge of reactions producing or destroying 
$^{92}$Nb in order to reduce as much as possible the nuclear uncertainties of astrophysical models. Namely, the rates of reactions 
(calculated from the cross section) must be known at the relevant stellar temperatures. In ref. \cite{Travaglio2011} (see their 
figure 5) it has been shown that most of the production of $^{92}$Nb takes place at a temperature around $T= 2.5 - 2.7$ GK, where 
$^{20}$Ne burning occurs.

The production of $^{92}$Nb is governed by the destruction of $^{93}$Nb and $^{92}$Zr seeds. It also gets some indirect contributions from $^{91,94,96}$Zr via $^{92}$Zr. The nuclide $^{92}$Nb is mainly destroyed by the reaction $^{92}$Nb($\gamma$,n)$^{91}$Nb, while three reactions produce it, $^{93}$Nb($\gamma$,n)$^{92}$Nb, $^{92}$Zr(p,n)$^{92}$Nb and  $^{91}$Zr(p,$\gamma$)$^{92}$Nb. 

While the experimental study of $\gamma$-induced reactions is technically very hard (and does not even provide the necessary astrophysical reaction rate, see e.g. \cite{Mohr2007}), proton-induced reactions can be measured more easily. The measured cross section of $^{92}$Zr(p,n)$^{92}$Nb is available in the literature \cite{Flynn1979}, for the 
$^{91}$Zr(p,$\gamma$)$^{92}$Nb, however, there is no experimental data at all.  Therefore, the aim of the present work was to 
measure this cross section in the energy range (Gamow-window \cite{PhysRevC.81.045807}) relevant for the p-process nucleosynthesis. 
The Gamow-window for this reaction at temperatures $T=2.5-2.7$ GK cited above lies between about 1.5 and 2.8\,MeV.

In the same experiment we were also able to determine the cross sections of $^{96}$Zr(p,n)$^{96}$Nb, see below for further details.

\section{\label{sec:exp}Experimental procedure}
\subsection{\label{subsec:invesreac}Investigated reactions}

As outlined above, the primary aim of the present work was the study of the $^{91}$Zr(p,$\gamma$)$^{92m}$Nb reaction using the activation method \cite{Gyurky2019}. The cross section was determined based on the off-line detection of the $\gamma$-radiation following the $\beta$-decay of the reaction product. The natural isotopic abundance of $^{91}$Zr is relatively high (11.22\,\%), therefore isotopically enriched material was not inevitable. Natural isotopic composition target material was thus used which in principle allows the study of proton induced reactions on other naturally occurring Zr isotopes. 

Zr has five stable isotopes with mass numbers 90, 91, 92, 94 and 96. Proton induced reactions on these isotopes often lead to radioactive Nb isotopes. However, taking into account the half-lives of the reaction products and the relative intensities of the emitted $\gamma$-radiation, the only other reaction channel which could be measured was $^{96}$Zr(p,n)$^{96}$Nb. Hence, in the present work the cross sections of $^{91}$Zr(p,$\gamma$)$^{92m}$Nb and $^{96}$Zr(p,n)$^{96}$Nb are presented. Table\,\ref{tab:decaydata} shows the relavant decay parameters of the reaction products populated in these reactions.

\begin{table}
\caption{\label{tab:decaydata} Decay data of the reaction products. Only those $\gamma$-transitions are listed which have been used for the analysis. The data are taken from \cite{BAGLIN20122187,ABRIOLA20082501}.}
\begin{indented}
\item[]\begin{tabular}{lccc}
\br
Isotope & Half-life & $\gamma$-energy & Relative \\
 				&						& [keV] 					& $\gamma$-intensity [\%] \\
\mr
$^{92m}$Nb & 10.15\,$\pm$\,0.02\,d & 934.4 & 99.15\,$\pm$\,0.04\\
$^{96}$Nb & 23.35\,$\pm$\,0.05\,h & 460.0 & 26.62\,$\pm$\,0.19\\
						&											& 568.9 & 58.0\,$\pm$\,0.3\\
						&											& 778.2 & 96.45\,$\pm$\,0.22\\
						&											& 1200.2 & 19.97\,$\pm$\,0.10\\

\br
\end{tabular}
\end{indented}
\end{table}

It is worth noting that in the case of the $^{91}$Zr(p,$\gamma$)$^{92}$Nb reaction only the partial cross section section leading to the isomeric state of $^{92}$Nb (denoted as $^{92m}$Nb) could be measured as the ground state has too long half-life of 3.47$\cdot 10^7$ years. However, as it will be shown in section\,\ref{sec:results} the total cross section is dominated by the measured partial one.

\subsection{\label{subsec:target}Target preparation and characterization}

Thin Zr targets were prepared by electron beam evaporation of natural isotopic composition metallic Zr onto 6.5\,$\mu$m thick Al foil backings. The first information about the target thickness was obtained by weighing, the weight of the Al foil was measured with 1\,$\mu$g accuracy before and after the evaporation. It was however observed that during or after the evaporation, oxidation of the Zr layer occurs, and thus weighing does not provide precise information about the effective target thickness. 

\begin{figure}
\centering
\includegraphics[width=0.49\textwidth]{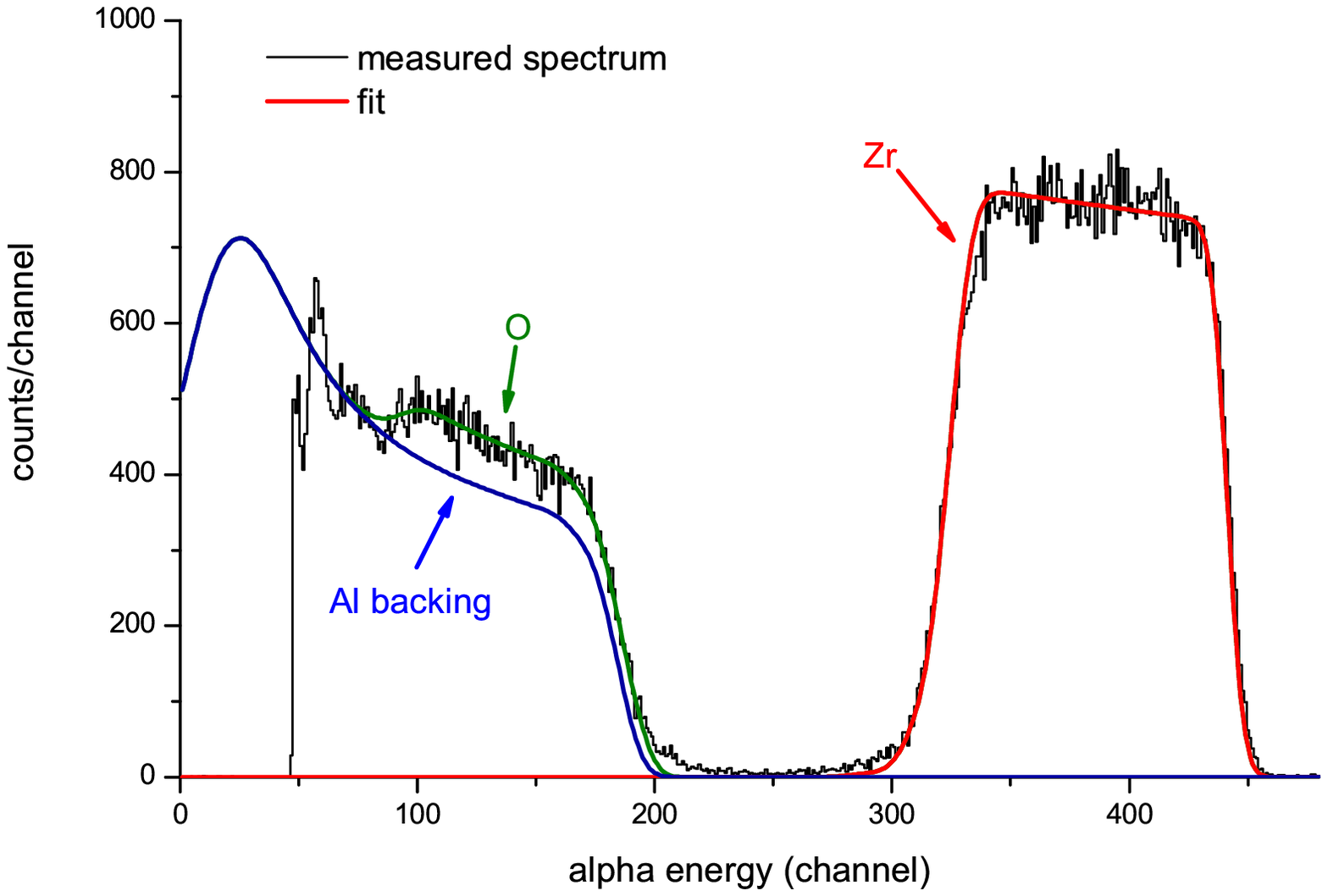}
\includegraphics[width=0.49\textwidth]{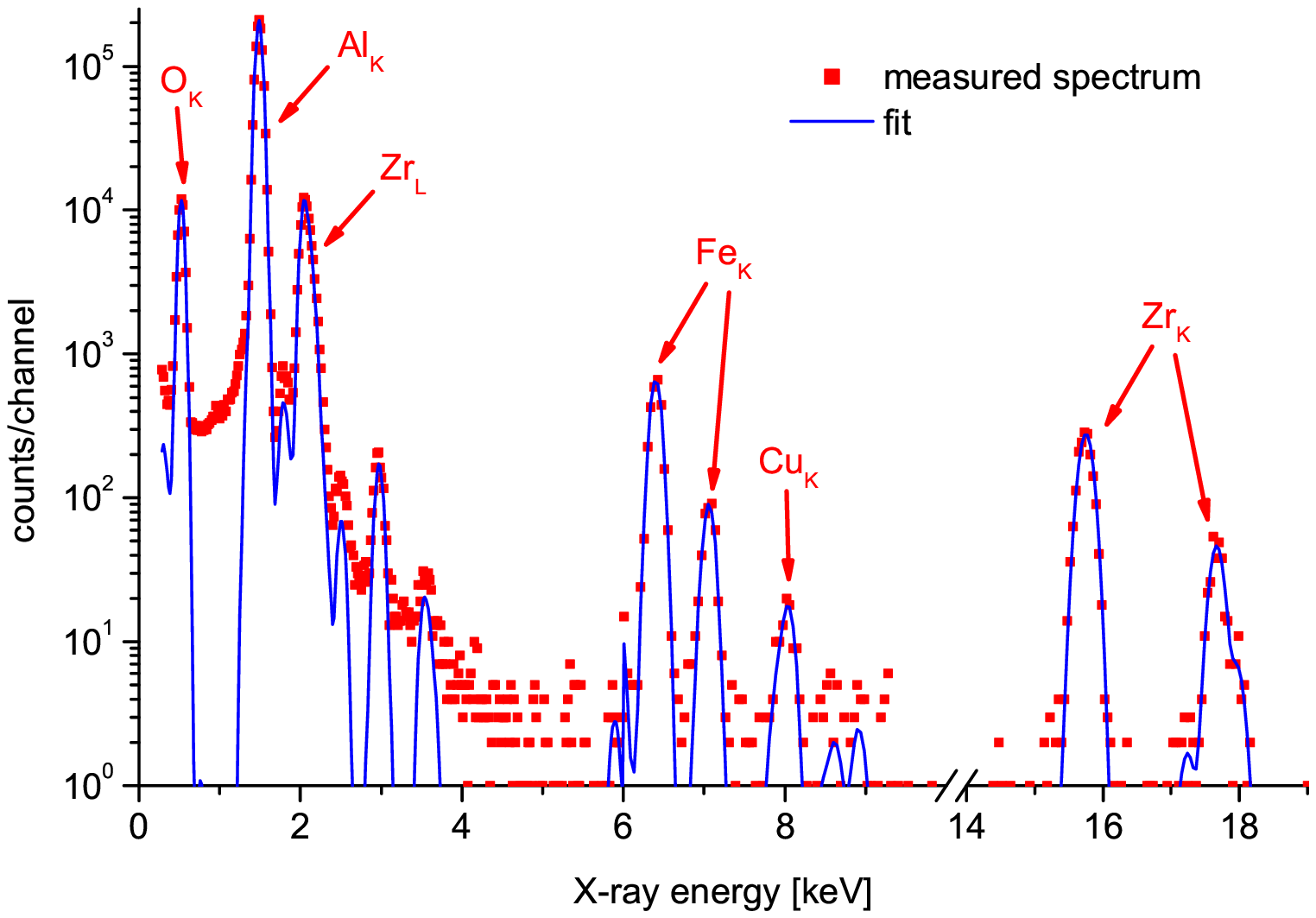}
\caption{\label{fig:PIXE_RBS} RBS (left panel) and PIXE (right panel) spectra measured for the target thickness determination. Fits 
of the measured data are also shown and the various components of the targets are indicated.}
\end{figure}

Two other techniques were therefore applied to determine the number of Zr atoms in the target, which is the relevant quantity for the cross section calculation. Rutherford Backscattering Spectrometry (RBS) and Particle Induced X-ray Emission (PIXE) methods were used. Both experiments were carried out at the microbeam setup on the 5\,MV Van de Graaff accelerator of ATOMKI \cite{RAJTA1996148}. Further details of the setup and the measurements can be found in \cite{Gyurky2019,Kiss_2021,Huszank_RBS}. Shortly, the RBS measurement utilized a 1.6 or 2.0\,MeV $\alpha$-beam and two particle detectors placed at backward angles of 135 and 165 degrees. The obtained spectra were analyzed with the SIMNRA code \cite{SIMNRA}. For the PIXE measurement the targets were bombarded by 2.0\,MeV protons and the induced x-rays were detected by two detectors, a silicon drift x-ray detector (SDD) and a Gresham type Be windowed Si(Li) X-ray detector \cite{Kertesz2015}. The PIXE spectra were fitted using the GUPIX code \cite{CAMPBELL20103356} in order to obtain the Zr target thickness. Figure \ref{fig:PIXE_RBS} shows two typical spectra measured with the RBS (left panel) and PIXE methods (right panel). 

The total (statistical and systematic) uncertainty of the RBS measurements was about 6\,\%. Due to some technical problems the uncertainty of the PIXE measurement was larger, roughly 10\,\%. In the case of targets where both techniques were used, the results agreed within the uncertainty of the two methods. The final target thicknesses and their uncertainties were obtained as the weighted average of the two methods, which, owing to the difference in uncertainty, were dominated by the RBS result. 

Altogether 16 targets were prepared and analyzed and 7 of them were used for the cross section measurements in the present 
work\footnote{Some other Zr targets were used in a recent experiment on the $^{96}$Zr($\alpha$,n)$^{99}$Mo reaction 
\cite{Kiss_2021}.}. Their thicknesses range between (0.8\,--\,1.5)$\cdot 10^{18}$ Zr atoms/cm$^2$ with an uncertainty of 5\,\%. 

\subsection{\label{subsec:irrad}Irradiation}

The proton beam for the irradiation was provided by the Tandetron accelerator of ATOMKI \cite{2021EPJP..136..247B}. Based on a 
recent calibration \cite{RAJTA2018125}, the exact energy of the proton beam is known to a precision of better than 1\,keV. The cross 
section was measured in the proton energy range between 1450 and 2800\,keV covering completely the Gamow window for $^{91}$Zr(p,$\gamma$)$^{92}$Nb at the relevant temperatures of the p-process. The target chamber described e.g. in ref 
\cite{PhysRevC.91.034610} was used. The chamber served as a Faraday cup allowing the number of projectiles impinging on the target 
to be determined based on charge measurement. The typical beam intensity was about 4-5\,$\mu$A. The charge collected on the target 
was integrated and recorded in multichannel scaling mode with 1 min dwell time in order to take into account any variation in the 
beam intensity during the activation. The total lengths of the irradiations varied between 5 and 48 hours. 

In most cases two targets, placed behind each other were irradiated in a single activation in order to reduce the beam time requirement (hereafter referred to as front and rear targets). The Zr layer and the 6.5\,$\mu$m thick Al foil backing of the front targets caused energy losses in the range of about 100-200\,keV before the beam reached the rear target. See section \ref{sec:results} for the discussion about the related energy uncertainty.  

\subsection{\label{subsec:detection}Detection of the $\gamma$-radiation}

After the irradiation the targets were removed from the chamber and taken to the off-line counting setup. The $\gamma$-radiation following the $\beta$-decay of the reaction products was measured with a 100\,\% relative efficiency HPGe detector placed in a complete 4$\pi$ lead shielding of 10\,cm thickness against laboratory background radiation. 

Owing to the large half-life difference of the two studied reaction products ($^{92m}$Nb and $^{96}$Nb, see table\,\ref{tab:decaydata}) the $\gamma$-counting of each target was separated into two periods. About one hour after the irradiation the counting of one target started. Typically this was the rear target as in this case the lower beam energy results in a lower cross section and hence lower activity. This first measurement was carried out for about 24 hours. Then the front target was counted, again for at least 24 hours. In this first period the decay of $^{96}$Nb was measured for the $^{96}$Zr(p,n)$^{96}$Nb cross section determination. 

\begin{figure}
\centering
\includegraphics[width=0.49\textwidth]{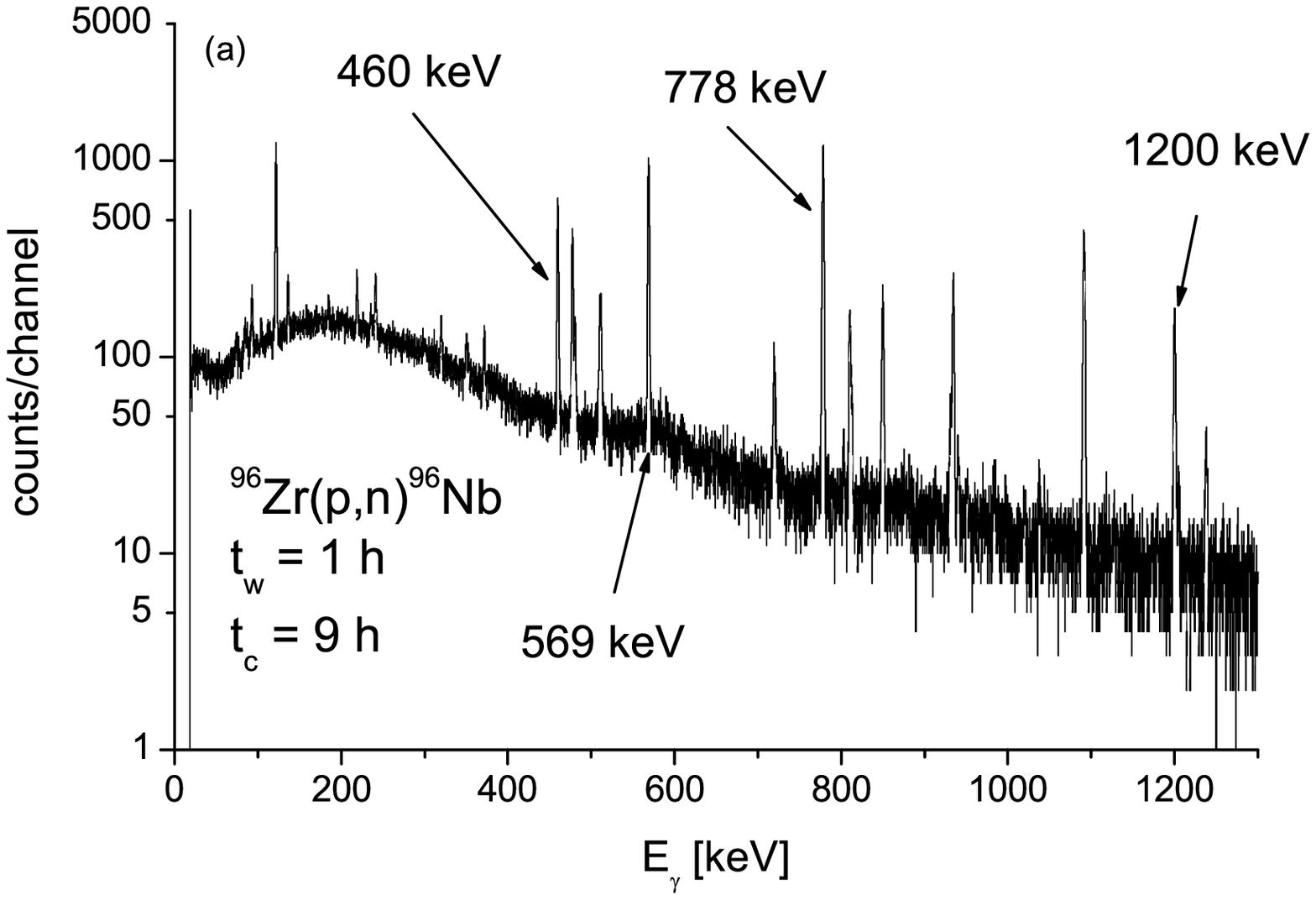}
\includegraphics[width=0.49\textwidth]{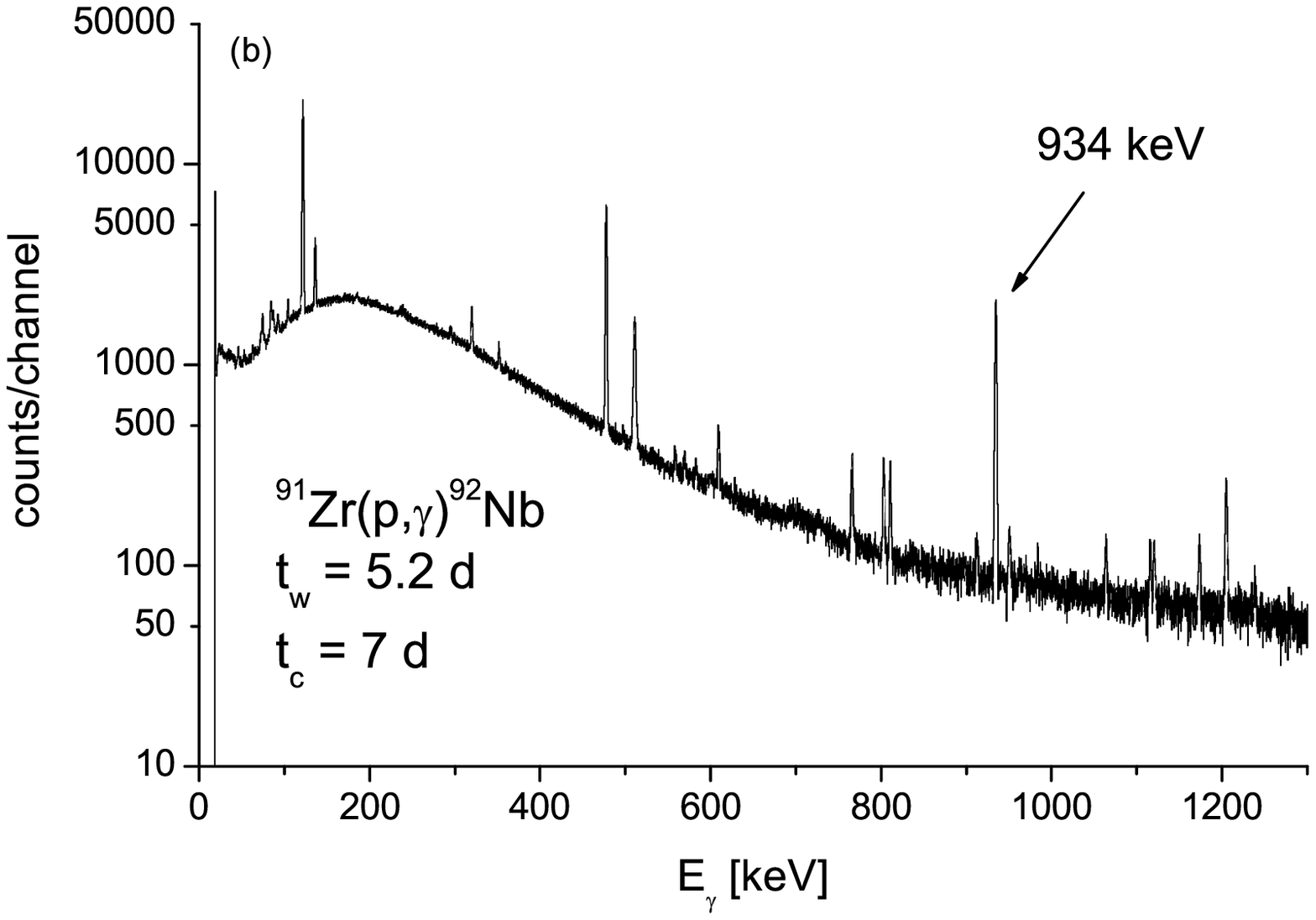}
\caption{\label{fig:spectrum} Typical $\gamma$-spectra measured 1 hour (left panel) and 5.2 days (right panel) after irradiation for 9 hours and 7 days, respectively. The left spectrum was used for the $^{96}$Zr(p,n)$^{96}$Nb cross section evaluation while the right one for the $^{91}$Zr(p,$\gamma$)$^{92m}$Nb reaction. The $\gamma$-peaks used for the analysis are indicated.}
\end{figure}

After these short counting periods, both targets were measured again, now for several days. Depending on the activity of the samples, this second counting lasted between 4 and 30 days. In this period the decay of $^{92m}$Nb was measured for the $^{91}$Zr(p,$\gamma$)$^{92m}$Nb cross section determination. Typical $\gamma$-spectra recorded in the first and second counting periods are shown in figure\,\ref{fig:spectrum} where the peaks used for the cross section determination are indicated, corresponding to transitions listed in table\,\ref{tab:decaydata}.

In order to obtain cross sections as low as possible at astrophysical energies, low target activities had to be measured. For this reason the detection efficiency was maximized by using a close counting geometry. The targets were placed 1\,cm far from the entrance window of the detector end cap.  In such a close geometry the true coincidence summing effect makes the precise absolute efficiency determination difficult. The two distances method, as described e.g. in ref \cite{Gyurky2019} was therefore used. The detection efficiency was measured with calibrated radioactive sources at far geometry, where the summing effect is negligible. A strong source containing $^{92m}$Nb and $^{96}$Nb was then prepared by irradiating a Zr target with a high energy (E$_p$\,=\,4\,MeV) proton beam. This strong source was measured at both far and close geometries and efficiency conversion factors between the two geometries were determined for all studied transitions. The uncertainty of the detection efficiency was 5\,\% which contains the uncertainty of the efficiency measured at far geometry and the uncertainty of the conversion factors.

\section{\label{sec:results}Experimental results}

The cross sections of the two studied reactions were measured in the proton energy range between E$_p$\,=\,1.45 and 2.80\,MeV. The lowest studied energy was determined by the strongly decreasing cross section. At E$_p$\,=\,1.45\,MeV only the  $^{91}$Zr(p,$\gamma$)$^{92m}$Nb cross section could be obtained, the yield from the $^{96}$Zr(p,n)$^{96}$Nb reaction was too low. At E$_p$\,=\,2.82\,MeV the $^{92}$Zr(p,n)$^{92}$Nb reaction channel opens which leads to the production of the same residual nucleus as $^{91}$Zr(p,$\gamma$)$^{92}$Nb. With the activation technique these two reactions cannot be distinguished and therefore the studied energy range was limited below this reaction threshold.

Table\,\ref{tab:results} shows the measured cross sections for the two reactions. In the first column the primary proton energies from the Tandetron accelerator are listed. Asterisks show those cases where the cross section was determined on a rear target (see section\,\ref{subsec:irrad}). The second and third columns contain the effective center-of-mass (c.m.) energy and its uncertainty for the two reactions. The energy uncertainty does not correspond to the whole energy range covered by the beam in the target. Instead it was calculated by taking into account the uncertainties of the primary beam energy, the target thickness and the stopping power values provided by the SRIM code \cite{SRIM}. In the case of rear targets the energy loss of the beam in the front targets and its backing was also taken into account for the uncertainty calculations. 

In order to increase the reliability of the measured cross sections, several measurements at the same energy were repeated with 
different targets. The results were always consistent within the uncertainties, as it can be seen in the table by checking the cross section values in adjacent rows corresponding to the same primary beam energy. Another reason for 
the repeated energies was the behavior of the $^{91}$Zr(p,$\gamma$)$^{92m}$Nb cross section. At the lowest energies the points show 
some fluctuation, they do not follow a smooth curve. The repeated measurements prove that this is not an experimental error but 
indeed the characteristics of the excitation function. The $^{96}$Zr(p,n)$^{96}$Nb cross section points do not exhibit such a 
fluctuation.

\begin{table}
\caption{\label{tab:results} Measured cross section of the $^{91}$Zr(p,$\gamma$)$^{92m}$Nb and $^{96}$Zr(p,n)$^{96}$Nb reactions.}
\begin{indented}
\item[]\begin{tabular}{lr@{\hspace{1mm}}c@{\hspace{1mm}}lr@{\hspace{1mm}}c@{\hspace{1mm}}l@{\extracolsep{2mm}}r@{\hspace{1mm}}c@{\hspace{1mm}}lr@{\hspace{1mm}}c@{\hspace{1mm}}l}
\br
E$_{\rm beam}$ & \multicolumn{6}{c}{E$^{\rm eff}_{\rm c.m.}$ [keV]} & \multicolumn{6}{c}{cross section [$\mu$barn]}\\
 \cline{2-7}
 \cline{8-13}
{[keV]}				&	 \multicolumn{3}{c}{$^{91}$Zr(p,$\gamma$)$^{92m}$Nb} & \multicolumn{3}{c}{$^{96}$Zr(p,n)$^{96}$Nb}				& \multicolumn{3}{c}{$^{91}$Zr(p,$\gamma$)$^{92m}$Nb} & \multicolumn{3}{c}{$^{96}$Zr(p,n)$^{96}$Nb}\\
\mr
1452.9	&	1434.3	&	$\pm$	&	2.2	&	1435.1	&	$\pm$	&	2.2	&	0.300	&	$\pm$	&	0.063	&		&		&		\\
1747.1$^*$	&	1493	&	$\pm$	&	23	&	1500	&	$\pm$	&	23	&	0.446	&	$\pm$	&	0.037	&	0.276	&	$\pm$	&	0.037	\\
1747.1$^*$	&	1493	&	$\pm$	&	23	&	1500	&	$\pm$	&	23	&	0.451	&	$\pm$	&	0.057	&	0.259	&	$\pm$	&	0.092	\\
1874.5$^*$	&	1631	&	$\pm$	&	22	&	1637	&	$\pm$	&	22	&	0.973	&	$\pm$	&	0.085	&	0.936	&	$\pm$	&	0.090	\\
1874.5$^*$	&	1631	&	$\pm$	&	22	&	1637	&	$\pm$	&	22	&	0.893	&	$\pm$	&	0.080	&	1.03	&	$\pm$	&	0.13	\\
1747.1	&	1725.1	&	$\pm$	&	2.2	&	1726.1	&	$\pm$	&	2.2	&	2.58	&	$\pm$	&	0.22	&	2.14	&	$\pm$	&	0.25	\\
1747.1	&	1725.4	&	$\pm$	&	2.1	&	1726.4	&	$\pm$	&	2.1	&	2.78	&	$\pm$	&	0.22	&	2.21	&	$\pm$	&	0.22	\\
1996.1$^*$	&	1760	&	$\pm$	&	21	&	1766	&	$\pm$	&	21	&	3.02	&	$\pm$	&	0.24	&	2.83	&	$\pm$	&	0.24	\\
1874.5	&	1851.3	&	$\pm$	&	2.2	&	1852.3	&	$\pm$	&	2.2	&	6.26	&	$\pm$	&	0.50	&	5.55	&	$\pm$	&	0.44	\\
2096.1$^*$	&	1867	&	$\pm$	&	21	&	1872	&	$\pm$	&	20	&	7.18	&	$\pm$	&	0.56	&	6.03	&	$\pm$	&	0.48	\\
1996.1	&	1972.0	&	$\pm$	&	2.1	&	1973.2	&	$\pm$	&	2.1	&	15.3	&	$\pm$	&	1.2	&	11.5	&	$\pm$	&	0.9	\\
2096.1	&	2070.7	&	$\pm$	&	2.1	&	2071.9	&	$\pm$	&	2.1	&	22.6	&	$\pm$	&	1.7	&	20.0	&	$\pm$	&	1.6	\\
2394.2$^*$	&	2180	&	$\pm$	&	19	&	2184	&	$\pm$	&	18	&	17.7	&	$\pm$	&	1.4	&	38.5	&	$\pm$	&	3.0	\\
2394.2	&	2365.8	&	$\pm$	&	2.1	&	2367.1	&	$\pm$	&	2.1	&	19.3	&	$\pm$	&	1.5	&	96.7	&	$\pm$	&	7.6	\\
2698.1$^*$	&	2495	&	$\pm$	&	17	&	2498	&	$\pm$	&	17	&	26.2	&	$\pm$	&	2.1	&	185	&	$\pm$	&	14	\\
2700.0$^*$	&	2496	&	$\pm$	&	17	&	2500	&	$\pm$	&	17	&	28.2	&	$\pm$	&	1.7	&	176	&	$\pm$	&	14	\\
2698.1	&	2666.6	&	$\pm$	&	2.1	&	2668.1	&	$\pm$	&	2.1	&	28.6	&	$\pm$	&	2.2	&	362	&	$\pm$	&	28	\\
2700.0	&	2668.7	&	$\pm$	&	2.1	&	2670.2	&	$\pm$	&	2.1	&	29.1	&	$\pm$	&	2.3	&	353	&	$\pm$	&	27	\\
2800.0	&	2767.7	&	$\pm$	&	2.1	&	2769.3	&	$\pm$	&	2.1	&	32.5	&	$\pm$	&	2.6	&	502	&	$\pm$	&	39	\\

\br
\end{tabular}
\end{indented}
\end{table}

\section{\label{sec:discussion}Discussion}

\begin{figure}
\centering
\includegraphics[width=0.7\textwidth,angle=-90]{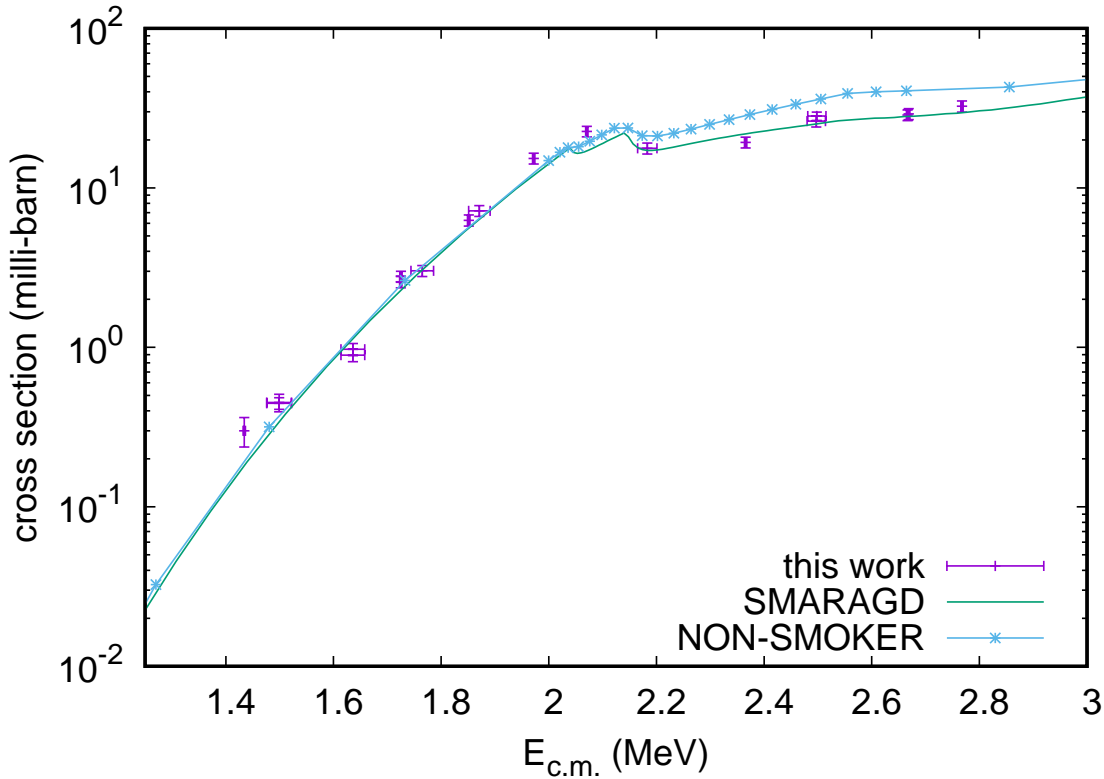}
\caption{\label{fig:zr91xs} Comparison of the experimental $^{91}$Zr(p,$\gamma$)$^{92m}$Nb cross sections with theoretical values 
obtained with the default settings of the NON-SMOKER and SMARAGD codes for $^{91}$Zr(p,$\gamma$)$^{92}$Nb. As detailed in the text, the theoretical curves are for the total, while the experimental data for the isomeric cross sections.}
\end{figure}

Figure \ref{fig:zr91xs} shows a comparison of the experimental cross sections for $^{91}$Zr(p,$\gamma$)$^{92m}$Nb and 
theoretical reaction cross sections for the reaction  $^{91}$Zr(p,$\gamma$)$^{92}$Nb. The data and the calculations clearly show the 
competition cusp between 2 MeV and 2.2 MeV which is caused by the opening of the (p,n) channel. The calculation with the default settings of the SMARAGD 
code \cite{smaragd,raurev} is in excellent agreement across the whole measured energy range. The default prediction of the older 
NON-SMOKER code \cite{nonsmoker,adndt} is in excellent agreement with the data below the opening of the neutron channel and 
slightly overestimates the cross sections above the channel opening. As shown in \cite{sensi}, the cross section is sensitive only 
to the proton width below the channel opening because the proton width is smaller than the $\gamma$ width in that energy region. On 
the other hand, the cross section is sensitive to neutron, proton, and $\gamma$ width above the neutron threshold. Assuming that 
the proton and neutron widths are predicted well (the neutron width is computed with a similar approach as the proton width), test 
calculations showed that a similar agreement as obtained with the SMARAGD code can be achieved by reducing the 
$\gamma$ width in the NON-SMOKER calculation.

Both predictions are for the reaction including transitions to the ground state and excited states in $^{92}$Nb, whereas the 
experiment determines the cross section by following the decay of the isomer $^{92m}$Nb. In a cascade calculation with the 
SMARAGD code, the $\gamma$-cascades during the de-excitation of populated excited states of $^{92}$Nb were followed and added up. 
It was found that 98\% of the de-excitations proceed through the isomer $^{92m}$Nb. This means that the predicted cross sections 
shown in figure\ \ref{fig:zr91xs} have to be reduced by only 2\% for a direct comparison with the data. On the scale of figure\ 
\ref{fig:zr91xs}, this is not visible because it would be about the width of a line.

Most astrophysical simulations make use of the nuclear reaction rates published in \cite{adndt,adndtxs}, which are based on the 
NON-SMOKER calculations. Computing the reaction rate for $^{91}$Zr(p,$\gamma$)$^{92}$Nb with the SMARAGD values, a reduction of the 
rate by a factor $0.9-0.93$ in the temperature range relevant for nucleosynthesis of the light $p$ nuclides is found. Due to the 
weak dependence of the $^{91}$Zr abundance on the $^{91}$Zr(p,$\gamma$) reaction, such a small change will not affect the 
astrophysical results of \cite{Travaglio2014}. It is possible to directly constrain the astrophysical rate by this measurement 
because not only the contribution of the $^{91}$Zr(p,$\gamma$)$^{92m}$Nb cross section to the total proton capture cross section 
of $^{91}$Zr is large but also the stellar reaction rate is completely determined by reactions on the ground state of $^{91}$Zr 
\cite{sensi}, i.e., the ground-state contribution to the stellar rate is 100\% for temperatures achieved during 
nucleosynthesis of the $p$ nuclides \cite{xfact}.

\begin{figure}
\centering
\includegraphics[width=0.7\textwidth,angle=-90]{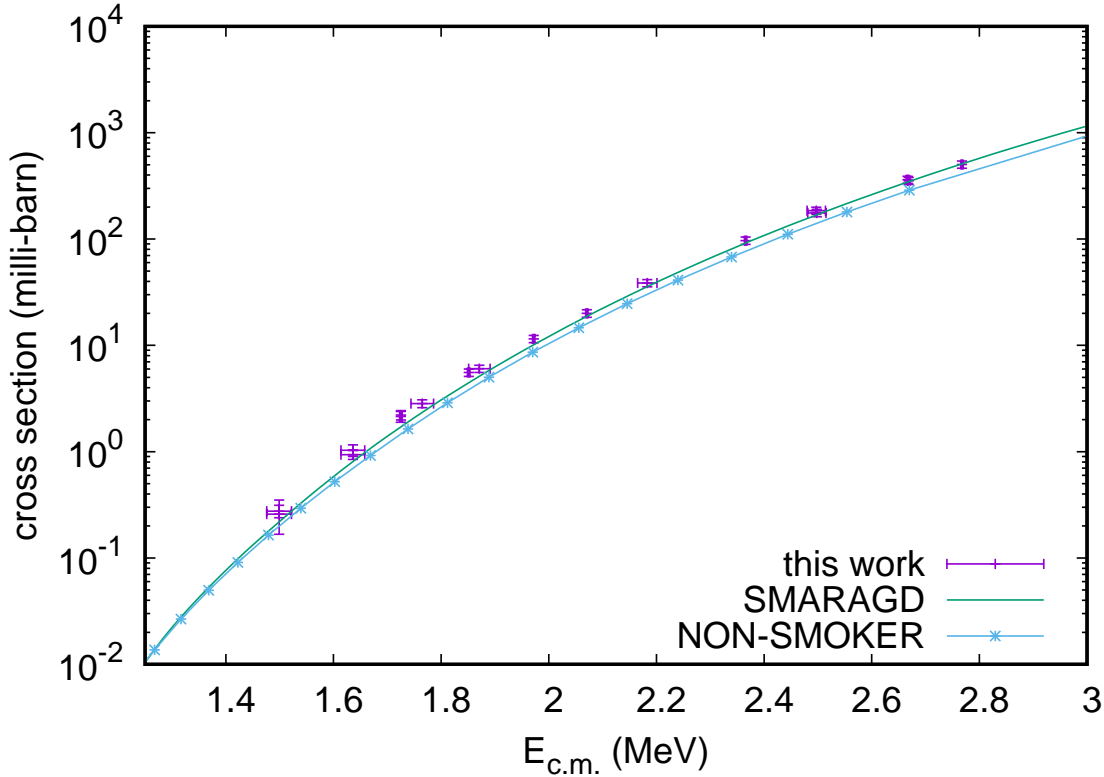}
\caption{\label{fig:zr96xs} Comparison of the experimental $^{96}$Zr(p,n)$^{96}$Nb cross sections with theoretical values 
obtained with the default settings of the NON-SMOKER and SMARAGD codes.}
\end{figure}

Figure \ref{fig:zr96xs} shows a comparison of experimental and theoretical cross sections for $^{96}$Zr(p,n)$^{96}$Nb\footnote{As opposed to the $^{91}$Zr(p,$\gamma$)$^{92}$Nb case, there are experimental cross section data for  $^{96}$Zr(p,n)$^{96}$Nb in the literature. Most of the measurements were carried out at higher energies, only the data of B.D. Kern et al. \cite{Kern1984} overlaps in energy with the present work. However, B.D. Kern et al. measured only partial cross sections to several states of the final nucleus and they did not report the total cross section. Therefore, the direct comparison with the present data is not possible.}. Both the 
NON-SMOKER and SMARAGD values are in good agreement with the data across the measured energy range, with the SMARAGD values being 
about 18\% larger and in slightly better agreement with the data. This reaction has not been identified as being of astrophysical 
interest but a confirmation of the predictions is interesting from a nuclear physics point-of-view, nonetheless.

Although the measured cross sections agree well with the previous prediction, here we also provide fits to the stellar 
reactivities derived from the SMARAGD calculation for use in astrophysical simulations. Not only do the SMARAGD calculations better 
reproduce the experimental data but also the fits are more accurate than the previous ones. The resulting fit parameters are given 
in Table \ref{tab:fitpars}. Note that reactivities should always be fitted in the direction of positive reaction $Q$-value to 
minimize numerical inaccuracies, therefore we give the parameters for $^{96}$Nb(n,p)$^{96}$Zr. The reactivity for the reverse 
reaction can be derived as explained in \cite{adndt}.


\begin{table}
\caption{\label{tab:fitpars}Fit parameters for the stellar reactivities of $^{91}$Zr(p,$\gamma$)$^{92}$Nb and 
$^{96}$Nb(n,p)$^{96}$Zr. The parameters follow the definition given in \cite{adndt}.}
\begin{center}
\begin{tabular}{cr@{\hspace{0mm}}lr@{\hspace{0mm}}l}
\br
&\multicolumn{2}{c}{$^{91}$Zr(p,$\gamma$)$^{92}$Nb}&\multicolumn{2}{c}{$^{96}$Nb(n,p)$^{96}$Zr}\\
\hline
$a_0$&$1.358905$&$\times 10^2$&$-4.180968$&$\times 10^1$\\
$a_1$&$-9.303907$&& $-3.610407$&$\times 10^{-1}$\\
$a_2$&$3.030056$&$\times 10^2$&  $2.039908$&$\times 10^1$\\
$a_3$&$-4.587240$&$\times 10^2$&  $2.314798$&$\times 10^1$\\
$a_4$&$2.254880$&$\times 10^1$&$-6.319860$&\\
$a_5$& $-1.198180$&&  $6.575093$&$\times 10^{-1}$\\
$a_6$&  $2.391198$&$\times 10^2$&  $1.254596$&$\times 10^1$\\
$a_0^\mathrm{rev}$&$1.586636$&$\times 10^2$& $-3.924473$&$\times 10^1$\\
\br
\end{tabular}
\end{center}
\end{table}

\section{Conclusions}

For the first time, we have measured the reaction cross sections of $^{91}$Zr(p,$\gamma$)$^{92m}$Nb and 
$^{96}$Zr(p,n)$^{96}$Nb. It was feasible to measure directly in the astrophysically relevant energy range by using the activation 
technique. Good agreement was found with the theoretical cross sections used in astrophysical models so far. Our results confirm 
and strengthen the astrophysical conclusions reached with those reaction models.

\ack
This work was supported by the European COST action "ChETEC" (CA16117), by NKFIH grants No. NN128072 and K134197 and by the \'UNKP-20-5-DE-2 and \'UNKP-20-5-DE-297 New National Excellence Programs of the Ministry of Human Capacities of Hungary. G.G. Kiss and T. Sz\"ucs acknowledge support from the Bolyai research fellowship of
the Hungarian Academy of Sciences. 

\section*{References}

\providecommand{\newblock}{}


\begin{thebibliography}{10}
\expandafter\ifx\csname url\endcsname\relax
  \def\url#1{{\tt #1}}\fi
\expandafter\ifx\csname urlprefix\endcsname\relax\def\urlprefix{URL }\fi
\providecommand{\eprint}[2][]{\url{#2}}

\bibitem{RevModPhys.83.157}
K\"appeler F, Gallino R, Bisterzo S and Aoki W 2011 {\em Rev. Mod. Phys.\/}
  {\bf 83}(1) 157--193
  \urlprefix\url{https://link.aps.org/doi/10.1103/RevModPhys.83.157}

\bibitem{RevModPhys.93.015002}
Cowan J~J, Sneden C, Lawler J~E, Aprahamian A, Wiescher M, Langanke K,
  Mart\'{\i}nez-Pinedo G and Thielemann F~K 2021 {\em Rev. Mod. Phys.\/} {\bf
  93}(1) 015002
  \urlprefix\url{https://link.aps.org/doi/10.1103/RevModPhys.93.015002}

\bibitem{ARNOULD20031}
Arnould M and Goriely S 2003 {\em Physics Reports\/} {\bf 384} 1--84 ISSN
  0370-1573
  \urlprefix\url{https://www.sciencedirect.com/science/article/pii/S0370157303002424}

\bibitem{Rauscher_2013}
Rauscher T, Dauphas N, Dillmann I, Fr{\"o}hlich C, F{\"u}l{\"o}p Z and
  Gy{\"u}rky G 2013 {\em Reports on Progress in Physics\/} {\bf 76} 066201
  \urlprefix\url{https://doi.org/10.1088/0034-4885/76/6/066201}

\bibitem{Travaglio2014}
Travaglio C, Gallino R, Rauscher T, Dauphas N, R{\"o}pke F~K and Hillebrandt W
  2014 {\em The Astrophysical Journal\/} {\bf 795} 141
  \urlprefix\url{https://doi.org/10.1088/0004-637x/795/2/141}

\bibitem{Travaglio_2018}
Travaglio C, Rauscher T, Heger A, Pignatari M and West C 2018 {\em The
  Astrophysical Journal\/} {\bf 854} 18
  \urlprefix\url{https://doi.org/10.3847/1538-4357/aaa4f7}

\bibitem{Habae2017750118}
Haba M~K, Lai Y~J, Wotzlaw J~F, Yamaguchi A, Lugaro M and Sch{\"o}nb{\"a}chler
  M 2021 {\em Proceedings of the National Academy of Sciences\/} {\bf 118} ISSN
  0027-8424 (\textit{Preprint}
  \eprint{https://www.pnas.org/content/118/8/e2017750118.full.pdf})
  \urlprefix\url{https://www.pnas.org/content/118/8/e2017750118}

\bibitem{Harper1996}
{Harper} Charles~L J 1996 {\em Astrophys. J.\/} {\bf 466} 437

\bibitem{Arlandini1999}
Arlandini C, Kappeler F, Wisshak K, Gallino R, Lugaro M, Busso M and Straniero
  O 1999 {\em The Astrophysical Journal\/} {\bf 525} 886--900
  \urlprefix\url{https://doi.org/10.1086/307938}

\bibitem{Travaglio1999}
Travaglio C, Galli D, Gallino R, Busso M, Ferrini F and Straniero O 1999 {\em
  The Astrophysical Journal\/} {\bf 521} 691--702
  \urlprefix\url{https://doi.org/10.1086/307571}

\bibitem{Dauphas2003}
{Dauphas} N, {Rauscher} T, {Marty} B and {Reisberg} L 2003 {\em Nucl. Phys.
  A\/} {\bf 719} C287--C295 (\textit{Preprint} \eprint{astro-ph/0211452})

\bibitem{Nishimura2018}
Nishimura(西村信哉) N, Rauscher T, Hirschi R, Murphy A~S~J, Cescutti G and
  Travaglio C 2017 {\em Monthly Notices of the Royal Astronomical Society\/}
  {\bf 474} 3133--3139 ISSN 0035-8711 (\textit{Preprint}
  \eprint{https://academic.oup.com/mnras/article-pdf/474/3/3133/22892024/stx3033.pdf})
  \urlprefix\url{https://doi.org/10.1093/mnras/stx3033}

\bibitem{Woosley1978}
{Woosley} S~E and {Howard} W~M 1978 {\em Astrophys. J. Suppl. Ser.\/} {\bf 36}
  285--304

\bibitem{Rauscher2002}
Rauscher T, Heger A, Hoffman R~D and Woosley S~E 2002 {\em The Astrophysical
  Journal\/} {\bf 576} 323--348 \urlprefix\url{https://doi.org/10.1086/341728}

\bibitem{Hayakawa2013}
Hayakawa T, Nakamura K, Kajino T, Chiba S, Iwamoto N, Cheoun M~K and Mathews
  G~J 2013 {\em The Astrophysical Journal\/} {\bf 779} L9
  \urlprefix\url{https://doi.org/10.1088/2041-8205/779/1/l9}

\bibitem{Lugaro2016}
Lugaro M, Pignatari M, Ott U, Zuber K, Travaglio C, Gy{\"u}rky G and
  F{\"u}l{\"o}p Z 2016 {\em Proceedings of the National Academy of Sciences\/}
  {\bf 113} 907--912 ISSN 0027-8424 (\textit{Preprint}
  \eprint{https://www.pnas.org/content/113/4/907.full.pdf})
  \urlprefix\url{https://www.pnas.org/content/113/4/907}

\bibitem{Travaglio2011}
Travaglio C, R{\"o}pke F~K, Gallino R and Hillebrandt W 2011 {\em The
  Astrophysical Journal\/} {\bf 739} 93
  \urlprefix\url{https://doi.org/10.1088/0004-637x/739/2/93}

\bibitem{Mohr2007}
Mohr P, F{\"u}l{\"o}p Z and Utsunomiya H 2007 {\em The European Physical
  Journal A\/} {\bf 32} 357--369 ISSN 1434-601X
  \urlprefix\url{https://doi.org/10.1140/epja/i2006-10378-y}

\bibitem{Flynn1979}
Flynn D~S, Hershberger R~L and Gabbard F 1979 {\em Phys. Rev. C\/} {\bf 20}(5)
  1700--1705 \urlprefix\url{https://link.aps.org/doi/10.1103/PhysRevC.20.1700}

\bibitem{PhysRevC.81.045807}
Rauscher T 2010 {\em Phys. Rev. C\/} {\bf 81}(4) 045807 \\
  \urlprefix\url{https://link.aps.org/doi/10.1103/PhysRevC.81.045807}

\bibitem{Gyurky2019}
Gy{\"u}rky G, F{\"u}l{\"o}p Z, K{\"a}ppeler F, Kiss G~G and Wallner A 2019 {\em
  The European Physical Journal A\/} {\bf 55} 41 ISSN 1434-601X
  \urlprefix\url{https://doi.org/10.1140/epja/i2019-12708-4}

\bibitem{BAGLIN20122187}
Baglin C~M 2012 {\em Nuclear Data Sheets\/} {\bf 113} 2187--2389 ISSN 0090-3752
  \urlprefix\url{https://www.sciencedirect.com/science/article/pii/S0090375212000671}

\bibitem{ABRIOLA20082501}
Abriola D and Sonzogni A 2008 {\em Nuclear Data Sheets\/} {\bf 109} 2501--2655
  ISSN 0090-3752
  \urlprefix\url{https://www.sciencedirect.com/science/article/pii/S0090375208000811}

\bibitem{RAJTA1996148}
Rajta I, Borbély-Kiss I, Mórik G, Bartha L, Koltay E and Kiss A~Z 1996 {\em
  Nuclear Instruments and Methods in Physics Research Section B: Beam
  Interactions with Materials and Atoms\/} {\bf 109-110} 148--153 ISSN
  0168-583X
  \urlprefix\url{https://www.sciencedirect.com/science/article/pii/0168583X95008977}

\bibitem{Kiss_2021}
Kiss G~G, Szegedi T~N, Mohr P, Jacobi M, Gy{\"u}rky G, Husz{\'{a}}nk R and
  Arcones A 2021 {\em The Astrophysical Journal\/} {\bf 908} 202
  \urlprefix\url{https://doi.org/10.3847/1538-4357/abd2bc}

\bibitem{Huszank_RBS}
Husz{\'a}nk R, Csedreki L, Kert{\'e}sz Z and T{\"o}r{\"o}k Z 2016 {\em Journal
  of Radioanalytical and Nuclear Chemistry\/} {\bf 307} 341

\bibitem{SIMNRA}
Mayer M Simnra version 6.06 sIMNRA version 6.06,
  https://www.max-planck-innovation.com/technology-offers/technology-offer/simnra-7-software.html

\bibitem{Kertesz2015}
Kert{\'e}sz Z, Furu E, Angyal A, Freiler {\'A}, T{\"o}r{\"o}k K and Horv{\'a}th
  {\'A} 2015 {\em Journal of Radioanalytical and Nuclear Chemistry\/} {\bf 306}
  283--288 ISSN 1588-2780
  \urlprefix\url{https://doi.org/10.1007/s10967-015-4175-5}

\bibitem{CAMPBELL20103356}
Campbell J, Boyd N, Grassi N, Bonnick P and Maxwell J 2010 {\em Nuclear
  Instruments and Methods in Physics Research Section B: Beam Interactions with
  Materials and Atoms\/} {\bf 268} 3356--3363 ISSN 0168-583X
  \urlprefix\url{https://www.sciencedirect.com/science/article/pii/S0168583X10006452}

\bibitem{2021EPJP..136..247B}
{Biri} S, {Vajda} I~K, {Hajdu} P, {R{\'a}cz} R, {Cs{\'\i}k} A, {Korm{\'a}ny} Z,
  {Perduk} Z, {Kocsis} F and {Rajta} I 2021 {\em European Physical Journal
  Plus\/} {\bf 136} 247

\bibitem{RAJTA2018125}
Rajta I, Vajda I, Gy{\"u}rky G, Csedreki L, Kiss A, Biri S, {van Oosterhout} H,
  Podaru N and Mous D 2018 {\em Nuclear Instruments and Methods in Physics
  Research Section A: Accelerators, Spectrometers, Detectors and Associated
  Equipment\/} {\bf 880} 125--130 ISSN 0168-9002
  \urlprefix\url{https://www.sciencedirect.com/science/article/pii/S0168900217311622}

\bibitem{PhysRevC.91.034610}
Yal\ifmmode \mbox{\c{c}}\else \c{c}\fi{}\ifmmode \imath~\else\i \fi{}n C,
  Gy{\"u}rky G, Rauscher T, Kiss G~G, {\"O}zkan N, G{\"u}ray R~T, Hal\'asz Z,
  Sz{\"u}cs T, F{\"u}l{\"o}p Z, Farkas J, Korkulu Z and Somorjai E 2015 {\em
  Phys. Rev. C\/} {\bf 91}(3) 034610
  \urlprefix\url{https://link.aps.org/doi/10.1103/PhysRevC.91.034610}

\bibitem{SRIM}
Ziegler J online, http://srim.org/ \,SRIM-2013 software code

\bibitem{smaragd}
Rauscher T 2014 code smaragd, v0.10.2 unpublished

\bibitem{raurev}
{Rauscher} T 2011 {\em International Journal of Modern Physics E\/} {\bf 20}
  1071--1169 (\textit{Preprint} \eprint{1010.4283})

\bibitem{nonsmoker}
{Rauscher} T and {Thielemann} F~K 1998 {Global statistical model calculations
  and the role of isospin} {\em Stellar Evolution, Stellar Explosions and
  Galactic Chemical Evolution\/} ed {Mezzacappa} A p 519 (\textit{Preprint}
  \eprint{nucl-th/9802040})

\bibitem{adndt}
{Rauscher} T and {Thielemann} F~K 2000 {\em Atomic Data and Nuclear Data
  Tables\/} {\bf 75} 1--351 (\textit{Preprint} \eprint{astro-ph/0004059})

\bibitem{sensi}
{Rauscher} T 2012 {\em The Astrophysical Journal\/} {\bf 201} 26
  (\textit{Preprint} \eprint{1205.0685})

\bibitem{adndtxs}
{Rauscher} T and {Thielemann} F~K 2001 {\em Atomic Data and Nuclear Data
  Tables\/} {\bf 79} 47--64 (\textit{Preprint} \eprint{nucl-th/0104003})

\bibitem{xfact}
{Rauscher} T 2012 {\em The Astrophysical Journal Letters\/} {\bf 755} L10
  (\textit{Preprint} \eprint{1207.1664})

\bibitem{Kern1984}
Kern B~D, F{\'e}nyes T, Krasznahorkay A, Dombr{\'a}di Z, Brant S and Paar V
  1984 {\em Nuclear Physics A\/} {\bf 430} 301--320 ISSN 0375-9474
  \urlprefix\url{https://www.sciencedirect.com/science/article/pii/0375947484900423}

\end{thebibliography}
\end{document}